\documentclass[12pt,a4paper]{article}
\usepackage{amssymb}

\def \Hspace {{\cal H}}
\def \Vspace {{\cal V}}
\def \Wspace {{\cal W}}
\def \AO {{\widehat A}}
\def \BO {{\widehat B}}
\def \CO {{\widehat C}}

\def \PO {{\widehat P}}

\def \rrangle {\rangle\!\rangle}
\def \llangle {\langle\!\langle}

\def \RSO {{\widehat {\mathbb R}}}

\def \PSO {{\check P}}
\def \PSV {|{\check P}\rangle\!\rangle}
\def \dangle {\rangle \langle }

\def \be {\begin{equation}}
\def \ee {\end{equation}}
\def \bea {\begin{eqnarray}}
\def \mea {\nonumber\\}
\def \eea {\end{eqnarray}}
\def \vs  {\vskip5mm}
\def \ni {\noindent} 
\begin{document}
\title{Entangled Subspaces and Quantum Symmetries}
\author{A.J. Bracken
\thanks{{\bf Email: ajb@maths.uq.edu.au}
It is a pleasure to thank G. Cassinelli 
and D. Ellinas for their generous hospitality 
at Universit\`a di Genova
and  TU Crete, respectively, where much of this work was completed.}
\\
Centre for Mathematical Physics\\ Department of Mathematics\\
University of Queensland\\
Brisbane 4072, Australia
}
\maketitle
\begin{abstract}
Entanglement is defined for each vector subspace of the tensor product of two finite-dimensional
Hilbert spaces, 
by applying the  notion of operator entanglement to the projection operator onto that
subspace.  The operator Schmidt decomposition of the projection operator defines a string of
Schmidt coefficients for each subspace, and this string is assumed to characterize its entanglement,
so that a first subspace is more entangled than a second, if the Schmidt string of the second
majorizes the Schmidt string of the first.  
The idea is applied to the antisymmetric and symmetric tensor products of a finite-dimensional
Hilbert space with itself, and also to the tensor product of an angular momentum $j$ with a spin
$1/2$.  When adapted to the subspaces  of states of the nonrelativistic hydrogen atom with
definite total angular momentum (orbital plus spin), within
the space of bound states with a given total energy, 
this leads to a complete ordering of those subspaces by their Schmidt strings.   
\end{abstract}

\section{Introduction}

If quantum entanglement \cite{horodecki} is to
be regarded as a physical resource \cite{nielsen1}, 
then it seems sensible to consider the entanglement not only of individual states, but
also of collections of states of a given composite quantum system. 
On the other hand, it is not clear how to combine measures of 
entanglement of individual states in such a
collection, because of the possibility of superposing given states to form new ones. 
Since the  natural organizational unit for 
any collection of states is the vector (sub)space spanned by
those states,  we are  led 
to the problem of quantifying
the degree of entanglement inherent
in a given vector subspace of the whole state space of a quantum system.
Examples of vector subspaces of interest  might be 
the space of
states with a 
given total energy, or  the space with a given total angular momentum.
More generally, a state subspace might be
labelled by the eigenvalues of any incomplete set of
commuting observables.  
A situation that arises often is one where the state space is associated with a tensor product
representation of some symmetry group or algebra, and the subspaces of interest carry  
irreducible subrepresentations of that algebra or group. Again, the example of angular momentum
springs to mind; we may be interested in a subspace of states carrying a definite total
angular momentum, for a quantum system made up of several subsystems, each contributing angular
momentum to the total.  
In such cases, the problem of quantifying the entanglement of  individual irreducible subspaces 
is
seen to have an  essentially
group-theoretical character; any measure of 
entanglement of such a subspace 
must surely involve such group-theoretical constructs as
the
Clebsch-Gordan coefficients of the corresponding group or algebra.
Conversely, considerations of the entanglement of such irreducible
subspaces seems likely to throw interesting new light on
familiar
group-theoretical reduction problems.

In what follows, we consider only bipartite systems, and vector subspaces $\Vspace$ of 
a complex, finite-dimensional state space
$\Hspace=\Hspace_1\otimes\Hspace_2$, where the two factor spaces have dimensions $d_1$ and $d_2$
respectively, and are equipped with the usual scalar products. 
Extensions to the multipartite case seem likely to
face the same sort of difficulties as entanglement
measures for state vectors of multipartite systems.  

\vs\ni
{\it Example 1\,}: 
As more specific motivation, consider the following 3-dimensional vector subspaces
of $\Hspace$,
in the case where $\Hspace_1\equiv \Hspace_2$ 
has orthonormal basis $\{e_1,\,e_2,\,e_3,\,\dots\,e_N\}$, $N\geq 3$:

\begin{itemize}
\item[(A)] 
$\Vspace_A$ is spanned by the  orthonormal vectors 
$(e_1\otimes e_2 -e_2\otimes e_1)/\sqrt{2}$,
\newline
$(e_2\otimes e_3 -e_3\otimes e_2)/\sqrt{2}$,
$(e_3\otimes e_1 -e_1\otimes e_3)/\sqrt{2}$.  

\item[(S)]
$\Vspace_S$ is spanned by the  orthonormal vectors 
$(e_1\otimes e_2 +e_2\otimes e_1)/\sqrt{2}$,
\newline
$(e_1\otimes e_1 -e_2\otimes e_2)/\sqrt{2}$,
$(e_1\otimes e_1 +e_2\otimes e_2)/\sqrt{2}$.
\end{itemize}
The bases of $\Vspace_A$ and $\Vspace_S$ 
so defined consist of three maximally entangled vectors in each
case.  However, it is 
possible to find another basis in $\Vspace_S$ with only one entangled vector, namely the set
$\{e_1\otimes e_2 + e_2\otimes e_1)/\sqrt{2}, e_1\otimes e_1, e_2\otimes e_2\}$, 
whereas 
every
choice of basis 
in $\Vspace_A$
consists entirely of maximally  entangled vectors.  
It is intuitive that 
$\Vspace_A$ is `more entangled' than $\Vspace_S$, and we seek to quantify such differences.

\section{Operator entanglement and subspaces}

In general, 
any vector subspace $\Vspace\leq \Hspace=\Hspace_1\otimes\Hspace_2$,  of 
dimension $1\leq d\leq d_1 d_2 $,  is 
characterized by a corresponding hermitian projection operator
$\PO$, 
\be
\PO\,\Vspace =\Vspace\,,\quad \PO^{\dagger}=\PO=\PO^2\,,\quad {\rm Tr}(\PO)=d\,,  
\label{projprops}
\ee
and we suggest that measures of entanglement of the operator $\PO$ provide suitable 
measures of entanglement of
$\Vspace$.  

Measures of operator entanglement have been considered previously in other contexts
\cite{nielsen2,zanardi}. The central idea is to consider each linear operator (matrix) $\AO$
as an `operator vector' $|A\rangle\!\rangle$ in the $(d_1d_2)^2$-dimensional `operator vector
space' $E_{\Hspace}$ of all linear operators on $\Hspace$,
with Hilbert-Schmidt scalar product
\be
\langle\!\langle A,B\rangle\!\rangle ={\rm Tr}(\AO^{\dagger} \BO)\,.
\label{hilbertschmidt}
\ee
Similarly, for $r=1,\,2$, linear operators on $\Hspace_r$ can be considered as
operator vectors in the $d_r^2$-dimensional operator vector space $E_{\Hspace_r}$ .
If $\AO=\BO\otimes\CO$, then 
$|A\rrangle=
|B\rrangle 
\otimes
|C\rrangle $,  
and $\AO$ is unentangled.
Otherwise, $\AO$ is entangled.

The operator vector 
\be
|{\check P}\rangle\!\rangle= \frac{1}{\sqrt{d}}\,|P\rangle\!\rangle 
\label{checkP}
\ee
 corresponding to the projector $\PO$ divided by $\sqrt{d}$,
is a unit operator
vector in $E_{\Hspace}$, according to (\ref{projprops}) and (\ref{hilbertschmidt}). 
We can define measures of entanglement of this unit
operator vector in $E_{\Hspace}$, 
just as we define measures of entanglement of unit vectors in $\Hspace$.  
To this end, we note firstly that $\PSV$ will have a Schmidt decomposition, 
\be
\PSV=
\sqrt{p_1}\, |E_1\rrangle\otimes |F_1\rrangle+
\sqrt{p_2}\, |E_2\rrangle\otimes |F_2\rrangle +\dots
\sqrt{p_K}\, |E_K\rrangle\otimes |F_K\rrangle\,,
\label{schmidtdecomp}
\ee
where 
\bea
K\leq {\bar K}={\rm min}\{d_1^2,\,d_2^2\}\,, \qquad\qquad
\mea
p_1\geq p_2\geq \dots \geq p_K>0\,,\quad p_1+p_2+\dots +p_K=1\,,  
\label{pprops}
\eea
while 
$|E_1\rrangle$, $|E_2\rrangle$, $\dots$ $|E_K\rrangle$ 
are orthonormal
operator vectors in $E_{\Hspace_1}$, and 
$|F_1\rrangle$, $|F_2\rrangle$, $\dots$ $|E_K\rrangle$ 
are orthonormal operator vectors in $E_{\Hspace_2}$.
If we introduce the superoperator \cite{caves} density matrix, of dimension
$(d_1d_2)^2\times(d_1d_2)^2$, 
\be
\RSO=|\PSO\rrangle\llangle \PSO|\,,
\label{superdensity1}
\ee
and then define the reduced superoperator density matrices  
$\RSO^{(1)}$ and
$\RSO^{(2)}$ by tracing over the second (respectively, the first) vector subspace of $E_{\Hspace}$,
then 
$|E_1\rrangle$, $|E_2\rrangle$, $\dots$ $|E_K\rrangle$ are eigen operator vectors of $\RSO^{(1)}$,
and 
$|F_1\rrangle$, $|F_2\rrangle$, $\dots$ $|E_K\rrangle$ 
are eigen operator vectors of $\RSO^{(2)}$, in each case
with
eigenvalues $p_1,\,p_2,\,\dots p_K$.   
Furthermore, the unentangled unit operator vector closest to $|\PSO\rrangle$ 
-- in the sense of the norm defined by
the scalar product (\ref{hilbertschmidt}) -- is
$|E_1\rrangle \otimes|F_1\rrangle$, and its distance from $|\PSO\rrangle$ is 
\bea
{\cal E}_D(|\PSO\rrangle)&=&\left((\llangle\PSO|-\llangle E_1|\otimes\llangle F_1|)
(|\PSO\rrangle - |E_1\rrangle\otimes |F_1\rrangle)\right)^{1/2} 
\mea
&=& 
\sqrt{2(1-\sqrt{p_1}\,)}\,.
\label{distance}
\eea
This distance provides a partial
measure of the entanglement of $|\PSO\rrangle$, and of $\Vspace$, and we shall
also write it as ${\cal E}_D(\Vspace)$.  

The entanglement of $\Vspace$ is fully characterized 
by its corresponding ${\bar K}$-dimensional `Schmidt string'
\be
{\cal S}({\Vspace})=
(p_1,\,p_2,\,\dots p_K,\,p_{K+1}=0,\,p_{K+2}=0,\,\dots,\,p_{\bar K}=0)\,.   
\label{schmidtvector}
\ee
Various partial measures of entanglement can be defined in terms of the Schmidt string, including
${\cal E}_D$ as above.  
Thus the `information' measure of entanglement of $|\PSO\rrangle$, and hence of $\Vspace$, is
\bea
{\cal E}_I(|\PSO\rrangle)&=&{\cal E}_I(\Vspace)
\mea
&=& -{\rm Tr}(\RSO^{(1)}\log_2(\RSO^{(1)})
= -{\rm Tr}(\RSO^{(2)}\log_2(\RSO^{(2)})
\mea
&=&-\sum_{\alpha=1}^K p_{\alpha}\log_2(p_{\alpha})\,,
\label{infoent}
\eea
while the `trace' measure of entanglement is
\bea
{\cal E}_T(|\PSO\rrangle)&=&{\cal E}_T(\Vspace)
\mea
&=&1-{\rm Tr}(\RSO^{(1)\,2})
=1-{\rm Tr}(\RSO^{(2)\,2})
\mea
&=& 1-\sum_{\alpha=1}^K p_{\alpha}^2\,.
\label{traceent}
\eea
A better indicator 
of entanglement is provided with the help of  the notion of {\it majorization}
\cite{marshall,wehrl,nielsen3}.  Thus we may say
that 
$\Vspace\leq \Hspace$ is more entangled than $\Wspace\leq \Hspace$, with
Schmidt string ${\cal S}({\Wspace})=(q_1,\,q_2,\,\dots 
)$, if 
$p_1\leq q_1$ AND $p_1+p_2\leq q_1+q_2$ AND $\dots$, that is to say,  
if ${\cal S}({\Vspace})$  is majorized  by   $ {\cal
S}({\Wspace})$, which we write as   
${\cal S}({\Vspace})\prec\, {\cal
S}({\Wspace})$. 
When 
${\cal S}({\Vspace})\prec\, {\cal
S}({\Wspace})$, it can be shown \cite{wehrl} that 
${\cal E}_D(\Vspace)\geq {\cal E}_D(\Wspace)$,    
${\cal E}_I(\Vspace)\geq {\cal E}_I(\Wspace)$     and 
${\cal E}_T(\Vspace)\geq {\cal E}_T(\Wspace)$.  But when neither of ${\cal S}({\Vspace})$ 
and ${\cal
S}({\Wspace})$
majorizes the other, some of these inequalities and not others may be reversed.  In that
situation, it is best to say only that $\Vspace$ and $\Wspace$ are {\it differently}
entangled.  

The preceding two paragraphs merely paraphrase for operator (or subspace) entanglement
what is well-known for state entanglement.
Many statements that hold true for the tensor product space of states $\Hspace$, go over to  the
tensor product space of operators $E_{\Hspace}$, without the need for new proofs.   
In what follows, we give some properties of subspace entanglement as defined, and then some 
examples of naturally arising entangled subspaces and their Schmidt
strings.   

\vs\ni
{\it Property 1\,}: If in the situation described above, $\Hspace_1$ is embedded as a
subspace in a
larger space $\Hspace_1'$, and $\Hspace_2$ is embedded as a subspace in
a larger space $\Hspace_2'$, 
so that
\be
\Vspace\leq\Hspace_1\otimes \Hspace_2 \leq\Hspace_1'\otimes \Hspace_2'\,,\quad
\Hspace_1\leq\Hspace_1'\,,
\quad \Hspace_2\leq\Hspace_2'\,,
\label{inclusions}
\ee
then ${\cal S} (\Vspace)$, when $\Vspace$ is regarded as a subspace of
$\Hspace_1'\otimes\Hspace_2'$, differs from ${\cal S}(\Vspace)$
when $\Vspace$ is regarded as a subspace of
$\Hspace_1\otimes \Hspace_2$, only by the addition of the appropriate
number of zeros on the right-hand
end.   In this sense, our notion of entanglement of a subspace is stable against embeddings.  

\vs\ni
{\it Property 2\,}: If $\Vspace$ has the form $\Vspace_1\otimes \Vspace_2$, where
$\Vspace_1\leq\Hspace_1$ and $\Vspace_2\leq\Hspace_2$, then 
$|\PSO\rrangle=|\PSO_1\rrangle\otimes |\PSO_2\rrangle$ is 
unentangled, and so also is $\Vspace$, according to our
definition.  
In this case, the Schmidt string has the form
\be
{\cal S}(\Vspace)=\Big(1,\,0,\,0,\,\dots,\,0\Big)\,.
\label{unentangledschmidt}
\ee
In particular, $\Hspace$ itself is unentangled, and ${\cal S}(\Hspace)$ has the form
(\ref{unentangledschmidt}).  

\vs\ni
{\it Property 3\,}: If $\Vspace$ is 1-dimensional, spanned by the unit vector $|v\rangle$ say,  
with Schmidt decomposition 
\be
|v\rangle 
=
\sum_{\alpha=1}^k\sqrt{p_{\alpha}}\,\,|e_{\alpha}\rangle\otimes|f_{\alpha}\rangle\,,
\label{vexpand}
\ee
then 
\be
\PO
=
|v\rangle\langle v|=\sum_{\alpha ,\,\beta=1}^k\sqrt{p_{\alpha}p_{\beta}}\,\,|e_{\alpha}\rangle\langle
e_{\beta}|\otimes
|f_{\alpha}\rangle\langle f_{\beta}|\,.
\label{vvexpand}
\ee
This defines the Schmidt decomposition of 
$|\PSO\rrangle$, whose Schmidt string then has as components the
$p_{\alpha}p_{\beta}$, with a suitable ordering.  
From this it is easily deduced that the Schmidt string of $|v\rangle$ majorizes the Schmidt string of
$|u\rangle$ if and only if the Schmidt string of the subspace spanned by $|v\rangle$
majorizes the Schmidt string of the subspace spanned by $|u\rangle$.   
This guarantees that
our notion of entanglement of 1-dimensional subspaces is consistent with that for state
vectors.  
In particular, it is also easily seen that 
\bea
{\cal E}_I(\Vspace)&=&-\sum_{\alpha,\,\beta=1}^k
p_{\alpha}p_{\beta}\log_2(p_{\alpha}p_{\beta})
\mea
&=& -2\sum_{\alpha=1}^kp_{\alpha}\log_2(p_{\alpha})=2{\cal E}_I(|v\rangle)\,.
\label{entrelation}
\eea
Thus the information measures of entanglement of a 1-dimensional subspace and of any unit vector
within that subspace differ only by the constant factor  2.

\section{Antisymmetric and symmetric subspaces}
\vs\ni
{\it Example 2\,}: As a  generalization of (A) in Example 1 above, consider  
the `antisymmetric tensor product' space  $\Vspace_A\leq
\Hspace=\Hspace_1\otimes\Hspace_2$, where
$\Hspace_1\cong \Hspace_2$ has an orthonormal basis
$\{|e_1\rangle,\,|e_2\rangle,\,\dots,\,|e_n\rangle \}$.  An orthonormal basis for $\Vspace_A$ is
provided by the $n(n-1)/2$ vectors 
\be
|e_{kl}\rangle=(|e_k\rangle\otimes|e_l\rangle 
-
|e_l\rangle\otimes|e_k\rangle )/\sqrt{2}\,,\quad k< l,\quad k,\,l\in\{ 1,\,2,\,\dots,\,n\},\, 
\label{antibasis}
\ee
and the projector onto $\Vspace_A$ is  then
\bea
\PO_A&=&\sum_{k<l=1}^n |e_{kl}\rangle\langle e_{kl}|\qquad\qquad
\mea
=\frac{1}{2} \sum_{k<l=1}^n&&\!\!\!\!\!\!\!\!\!\!\Big(    
|e_k\rangle \langle e_k |\otimes |e_l\rangle \langle e_l |
+
|e_l\rangle \langle e_l |\otimes |e_k\rangle \langle e_k |
\mea
&-&
|e_k\rangle \langle e_l |\otimes |e_l\rangle \langle e_k |
-
|e_l\rangle \langle e_k |\otimes |e_k\rangle \langle e_l |
\Big)\,. 
\label{projbasis1}
\eea
Labelling the unit operator vectors in $E_{\Hspace_1}$ and $E_{\Hspace_2}$ as 
\bea
|1\rrangle = |e_1\dangle e_1|,\, |2\rrangle =|e_2\dangle e_2 |,\,\dots,\, |n\rrangle =
|e_n\dangle e_n |,\,
\mea
|n+1\rrangle =|e_1\dangle e_2 |,\, |n+2\rrangle =|e_2\dangle e_1 |,\,\dots ,\,
|3n-2\rrangle=|e_n\dangle e_1|,\,
\mea
|3n-1\rrangle =|e_2\dangle e_3|,\,|3n\rrangle=|e_3\dangle e_2|,\,\dots,\,|5n-6\rrangle=|e_n\dangle
e_2|,
\mea
\dots ,\, |n^2\rrangle = |e_n\dangle e_{n-1} |\,,
\label{opbasis}
\eea
we then have from (\ref{projbasis1}), the unit operator vector
\be
|\PSO\rrangle =\frac{1}{\sqrt{2n(n-1)}}\sum_{r,s=1}^{n^2}{\cal A}_{rs}|r\rrangle\otimes |s\rrangle
\label{expansion1}
\ee
where the $n^2\times n^2$ matrix ${\cal A}$ with matrix elements ${\cal A}_{rs}$ takes the form
\be
{\cal A}={\cal B}
\oplus 
{\cal C}
\oplus 
{\cal C}
\oplus \dots \oplus 
{\cal C}\,.
\label{matrixform1}
\ee
Here ${\cal B}$ is $n\times n$, with all diagonal elements equal to $0$, and all nondiagonal
elements equal to $1$.  Each of the $n(n-1)/2$ copies of ${\cal C}$ is $2\times 2$, with diagonal
elements equal to $0$, and off-diagonal elements equal to $-1$.  

It follows from (\ref{expansion1})
that the matrix elements of $\RSO^{(1)}$ in this case are just those of ${\cal
A}{\cal A}^{\dag}$, whose eigenvalues are easily calculated from (\ref{matrixform1}) to be 
\be
{\cal S}(\Vspace_A)=\frac{1}{2n(n-1)}\Big((n-1)^2,\,1,\,1,\,\dots,\,1\Big)\,,
\label{antischmidtvector}
\ee
where the $1$ appears $n^2-1$ times.  
Then (\ref{antischmidtvector}) is the Schmidt string for $\Vspace_A$. It follows that
\bea
{\cal E}_D(\Vspace_A)=\sqrt{2(1-\sqrt{(n-1)/(2n)})},\quad\quad\qquad\qquad\qquad
\mea
{\cal E}_I(\Vspace_A)=\log_2\left(2n(n-1)^{1/n}\right),\,\,\,\,
{\cal E}_T(\Vspace_A)=(n+1)(3n-4)/[4n(n-1)]\,.
\label{antipartials}
\eea
Note that, as $n\to\infty$, ${\cal E}_D(\Vspace_A)$ and ${\cal E}_T(\Vspace_A)$ tend to constants,
whereas ${\cal E}_I(\Vspace_A)\sim \log_2 (n)$.   

\vs\ni
{\it Example 3}\,:
Consider again the space $\Hspace$ as in Example 2, and let $\Vspace_S$ denote the `symmetric
tensor product' space of dimension $n(n+1)/2$, 
with orthonormal basis $|e_{kl}\rangle$, $k\leq l= 1,\,2,\,\dots,\,n$, where 
\bea
|e_{kl}\rangle = (
|e_k\rangle \otimes |e_l\rangle +
|e_l\rangle \otimes |e_k\rangle 
)/\sqrt{2}\,, \quad k<l\,,
\mea
|e_{kk}\rangle = 
|e_k\rangle
\otimes
|e_k\rangle\,.\qquad\qquad\qquad
\label{symmbasis}
\eea
A similar calculation to that for the antisymmetric case shows that the Schmidt string for 
$\Vspace_S$ is 
\be
{\cal S}(\Vspace_S)= \frac{1}{2n(n+1)}\Big((n+1)^2,\,1,\,1,\,\dots,\,1\Big)\,,
\label{symmschmidtvector}
\ee
where again the $1$ appears $n^2-1$ times. Then  
\bea
{\cal E}_D(\Vspace_S)=\sqrt{2(1-\sqrt{(n+1)/(2n)})},\quad\quad\qquad\qquad\qquad
\mea
{\cal E}_I(\Vspace_S)=\log_2\left(2n/(n+1)^{1/n}\right),\,\,\,\,
{\cal E}_T(\Vspace_S)=(n-1)(3n+4)/[4n(n+1)]\,.
\label{symmpartials}
\eea
The asymptotic
behaviour of these quantities as $n\to\infty$ is similar to that in the antisymmetric case of
Example 2.  

We note from (\ref{antischmidtvector}) 
and (\ref{symmschmidtvector}) that, in Example 1,  the 3-dimensional antisymmetric and symmetric 
subspaces  of the $N^2$-dimensional space $\Hspace$ have $N^2$-dimensional
Schmidt strings 
\bea
{\cal S}(\Vspace_A)=\frac{1}{12}(4,\,1,\,1,\,1,\,1,\,1,\,1,\,1,\,1,\,0,\,0,\dots,\,0)
\mea
{\cal
S}(\Vspace_S)=\frac{1}{12}(9,\,1,\,1,\,1,\,0,\,0,\dots,\,0)\,,  
\label{ex1schmidts}
\eea
so that ${\cal S}(\Vspace_A)\prec\, 
{\cal S}(\Vspace_S)$, consistent with our intuition that $\Vspace_A$
is more entangled than $\Vspace_S$.

\section{Coupled angular momenta}
\vs\ni
{\it Example 4\,}: 
Consider the coupling of two  angular momenta,  
$(\widehat{J}_1,\,\widehat{J}_2,\,\widehat{J}_3)$ with spin $j$,
and $(\widehat{S}_1,\,\widehat{S}_2,\,\widehat{S}_3)$ with  spin $1/2$.
In this case, the full state space is   
\be
\Hspace=\Hspace_j\otimes\Hspace_{1/2}=\Vspace_{j+1/2}\oplus\Vspace_{j-1/2}\,,
\label{spinreduction}
\ee
where the space $\Hspace_k$, for $k=j$ or $1/2$, or $\Vspace_k$, for $k=j+1/2$ or $j-1/2$,
has dimension $(2k+1)$, and carries the corresponding 
irreducible representation of the spin
Lie algebra $su(2)$.
We are interested in the entanglement of $\Vspace_{j\pm 1/2}$, regarded as a subspace of $\Hspace$.  

Let 
$|m\rangle$, $m=j,\,j-1,\,\dots,\,-j$ denote the usual orthonormal basis of eigenstates of 
$\widehat{J}_3$ in
$\Hspace_j$, and let   
$|+\rangle$ and $|-\rangle$ denote the usual 
orthonormal basis of eigenstates of $\widehat{S}_3$
in $\Hspace_{1/2}$.
Then let 
\be
|1\rrangle=|+\rangle\langle+|\,,\quad
|2\rrangle=|-\rangle\langle-|\,,\quad
|3\rrangle=|+\rangle\langle-|\,,\quad
|4\rrangle=|-\rangle\langle+|\,,
\label{sopbasis1}
\ee
defining an orthonormal basis of operator vectors in $E_{1/2}$, and let 
\be
|m,n\rrangle=|m\rangle\langle n|\,,\quad m,\,n=j,\,j-1,\,\dots,\,-j\,,
\label{jopbasis1}
\ee
defining an orthonormal basis of operator vectors in $E_{j}$.  
Then $\widehat{S}_{\pm}=\widehat{S}_1\pm i\widehat{S}_2$ and $\widehat{S}_3$, regarded as operator
vectors in $E_{1/2}$, take the form 
\be
|S_+\rrangle = |3\rrangle\,,\quad|S_-\rrangle =|4\rrangle\,,\quad |S_3\rrangle=\frac{1}{2}(|1\rrangle
-|2\rrangle)\,.
\label{sopbasis2}
\ee
while $\widehat{J}_{\pm}$ and $\widehat{J}_3$, regarded as operator vectors in $E_j$,  take the form
\bea
|J_+\rrangle= 
\sqrt{(1)(2j)} |j,j-1\rrangle
+
\sqrt{(2)(2j-1)} |j-1,j-2\rrangle
+\dots 
\mea\mea 
\dots + \sqrt{(2j)(1)} |-j+1,-j\rrangle
\qquad\qquad\qquad
\mea\mea
|J_-\rrangle= 
\sqrt{(1)(2j)} |j-1,j\rrangle
+
\sqrt{(2)(2j-1)} |j-2,j-1\rrangle
+\dots 
\mea\mea
\dots +
\sqrt{(2j)(1)} |-j,-j+1\rrangle
\qquad\qquad\qquad
\mea\mea
|J_3\rrangle=
j|j,j\rrangle +
(j-1)|j-1,j-1\rrangle +\dots +
(-j)|-j,-j\rrangle\,.\qquad\qquad
\label{jopbasis2}
\eea
Recall \cite{edmonds} that the $\Hspace$ operator $\widehat{X}$, defined by
\be
\widehat{X}
=
\widehat{J}_+\otimes\widehat{S}_-
+\widehat{J}_-\otimes\widehat{S}_+
+2 \widehat{J}_3\otimes\widehat{S}_3\,,
\label{Xdef}
\ee 
takes the eigenvalue $j$ on the subspace
$\Vspace_{j+1/2}$ and the eigenvalue
$-(j+1)$ on the subspace $\Vspace_{j-1/2}$.  It follows that the projector from $\Hspace$  onto 
$\Vspace_{j\pm 1/2}$ is given by 
\be
\widehat{P}_{j\pm1/2}=\pm \frac{1}{2j+1}\Big
(\widehat{X}+\frac{1}{2}\widehat{I}\pm(j+\frac{1}{2})\widehat{I}\Big)\,,
\label{jprojectors}
\ee
where $\widehat{I}$ denotes the unit operator on $\Hspace$.  

Consider firstly the projector $\widehat{P}_{j+1/2}$. 
From (\ref{sopbasis2}), (\ref{jopbasis2}), (\ref{Xdef}) and (\ref{jprojectors}), 
we see that this operator, 
regarded as an operator vector on $E_{\Hspace}$, and normalized to a unit operator vector,
takes the form
\bea
|\check{P}_{j+1/2}\rrangle &=&
\frac{1}{\sqrt{2(j+1)}}\frac{1}{2j+1}
\Big\{
|J_+\rrangle \otimes|S_-\rrangle +|J_-\rrangle\otimes|S_+\rrangle 
\mea
\mea
&\quad&\qquad\qquad\qquad\qquad\qquad +2|J_3\rrangle\otimes|S_3\rrangle 
\mea
\mea
\!\!\!+ (j+1)\Big(|j,j\rrangle &+&|j-1,j-1\rrangle +\dots +
|-j,-j\rrangle\Big)\otimes\Big(|1\rrangle +|2\rrangle\Big)
\Big\}\,.
\mea
\label{plusprojdef}
\eea
Here the terms on the
last line represent the operator vector corresponding to the  operator $(j+1) \widehat{I}$.
Expression (\ref{plusprojdef}) has the general form
\be
|\check{P}_{j+1/2}\rrangle =
\sum_{(m,n)=(-j,-j)}^{(j,j)}\,\sum_{\alpha=1}^4
A_{(m,n),\alpha}|m,n\rrangle\otimes |\alpha\rrangle\,,
\label{Adef}
\ee
and we wish to claculate the eigenvalues of the reduced superoperator density matrix
\be
\RSO^{(2)}=
\sum_{(m,n)=(-j,-j)}^{(j,j)}\,\sum_{\alpha ,\beta=1}^4
\left\{A_{(m,n),\alpha}A_{(m,n),\beta}^{*}\right\}
|\alpha\rrangle \llangle\beta|\,,
\label{reducedR1}
\ee
or, what is the same thing, the eigenvalues of the $4\times 4$ matrix $Q$ with elements
\be
Q_{\alpha\beta}=
\sum_{(m,n)=(-j,-j)}^{(j,j)}\,
\left\{A_{(m,n),\alpha}A_{(m,n),\beta}^{*}\right\}\,.
\label{Qelements}
\ee
The only nonzero elements are, from (\ref{jopbasis2}) and (\ref{plusprojdef}), 
\bea
Q_{11}=Q_{22}&=& 
\frac{1}{2(j+1)}\frac{1}{(2j+1)^2}
\Big(
(2j+1)^2+(2j)^2+\dots + (1)^2
\Big)
\mea
\mea
&=&\frac{4j+3}{6(2j+1)}\,, 
\label{Q11}
\eea
\bea
Q_{33}=Q_{44}&=& 
\frac{1}{2(j+1)}\frac{1}{(2j+1)^2}
\Big(
(1)(2j)+(2)(2j-1)+\dots + (2j)(1)
\Big)
\mea
\mea
&=&\frac{j}{3(2j+1)}\,, 
\label{Q33}
\eea
and
\bea
Q_{12}&=&Q_{21}
\mea
&=& 
\frac{1}{2(j+1)}\frac{1}{(2j+1)^2}
\Big(
(2j+1)(1)+(2j)(2)+\dots + (1)(2j+1)
\Big)
\mea
\mea
&=&\frac{2j+3}{6(2j+1)}\,,
\label{Q12}
\eea
and the eigenvalues of $Q$ are now easily calculated to be $(j+1)/(2j+1)$ (multiplicity 1)
and $j/(6j+3)$ 
(multiplicity 3).
The Schmidt string of $\Vspace_{j+1/2}$ is therefore 
\be
{\cal S}(\Vspace_{j+1/2})=
\frac{1}{2j+1}\Big(j+1,\,\frac{j}{3},\,\frac{j}{3},\,\frac{j}{3}\Big)\,.
\label{jplusSchmidt}
\ee
We then have as scalar partial measures of the entanglement of this subspace,
\bea
{\cal E}_D(\Vspace_{j+1/2})=\sqrt{2(1-\sqrt{(j+1)/(2j+1)})},\quad\quad\qquad\qquad\qquad
\mea
{\cal E}_I(\Vspace_{j+1/2})=-\log_2\left(
\frac{\left(\frac{j}{3}\right)^{j/(2j+1)}(j+1)^{(j+1)/(2j+1)}}{2j+1}\right),
\mea
{\cal E}_T(\Vspace_{j+1/2})=2j(4j+3)/[3(2j+1)^2]\,.
\label{jpluspartials}
\eea
We see that as $j\to\infty$, 
all these quantities approach constant values.
This is a consequence of the fact that the Schmidt string (\ref{jplusSchmidt}) approaches the
constant value
\be
{\cal S}_0=\left(\frac{1}{2},\,
\frac{1}{6},\,
\frac{1}{6},\,
\frac{1}{6}\right)\,,
\label{limitingSchmidt}
\ee
and
can perhaps be understood as follows: as $j$ gets large, the number of states in $\Vspace_{j+1/2}$
with larger and larger 
positive or negative eigenvalue of 
$\widehat{J}_3\otimes \widehat{I}_{1/2}+
\widehat{I}_{j}\otimes \widehat{S}_{3}$ increases, and these states have smaller and
smaller entanglement, with the entanglement reaching zero for the highest and lowest weight
states.  

A similar calculation shows that the Schmidt string of $\Vspace_{j-1/2}$ is
\be
{\cal S}(\Vspace_{j-1/2})=
\frac{1}{2j+1}\Big(j,\,\frac{j+1}{3},\,\frac{j+1}{3},\,\frac{j+1}{3}\Big)\,.
\label{jminusSchmidt}
\ee
In this case, 
\bea
{\cal E}_D(\Vspace_{j-1/2})=\sqrt{2(1-\sqrt{j/(2j+1)})},\quad\quad\qquad\qquad\qquad
\mea
{\cal E}_I(\Vspace_{j-1/2})=-\log_2\left(
\frac{\left(\frac{j+1}{3}\right)^{(j+1)/(2j+1)}j^{j/(2j+1)}}{2j+1}\right),
\mea
{\cal E}_T(\Vspace_{j-1/2})=2(j+1)(4j+1)/[3(2j+1)^2]\,.
\label{jminuspartials}
\eea
As $j\to\infty$, these quantities approach the same constant values as in the previous case. 

Note that 
${\cal S}(\Vspace_{j+1/2})\succ\,
{\cal S}(\Vspace_{j-1/2})$, so that 
$\Vspace_{j-1/2}$
is more entangled than 
$\Vspace_{j+1/2}$.

\section{Application: Electron spin and  H-atom}

\vs\ni
{\it Example 5\,}: Consider the space $\Hspace^{(n)}$ of bound states of the nonrelativistic
hydrogen atom with principal quantum number $n$, where $n$ is a positive integer.
This space is  $n^2$-dimensional, with the structure
\be
\Hspace^{(n)}=\Hspace_0\oplus
\Hspace_1\oplus\dots\oplus
\Hspace_{n-1}\,,
\label{Hatomspace1}
\ee
where $\Hspace_l$ is $(2l+1)$-dimensional, 
corresponding to the orbital angular momentum content
$l=0,\,1,\,\dots,\,n-1$. Allowing for the spin of the electron, we have as the 
relevant state space including spin,
\bea
\Hspace&=&\Hspace^{(n)}\otimes\Hspace_{1/2}
\mea
&=&
\Big(\Hspace_0\otimes \Hspace_{1/2}\Big)\oplus
\Big(\Hspace_1\otimes \Hspace_{1/2}\Big)\oplus
\dots
\oplus
\Big(\Hspace_{n-1}\otimes \Hspace_{1/2}\Big)
\mea
&=&\Big({\cal V}_{1/2}\Big)\oplus
\Big({\widetilde {\cal  V}}_{1/2}\oplus{\cal V}_{3/2}\Big)\oplus
\Big({\widetilde {\cal V}}_{3/2}\oplus{\cal V}_{5/2}\Big)\oplus
\mea
&\quad&\dots\oplus
\Big({\widetilde{\cal V}}_{n-3/2}\oplus{\cal V}_{n-1/2}\Big)
\,.
\mea
\label{Hatomspace2}
\eea
From the results of Section 4, we see
that the Schmidt strings corresponding to these subspaces are 
\bea
{\cal S}({\cal V}_{k})
&=&\frac{1}{4k}\left(2k+1,\, 
\frac{2k-1}{3},\,
\frac{2k-1}{3},\,
\frac{2k-1}{3}
\right)
\mea
&\quad&\quad{\rm for}\quad
k=\frac{1}{2},\,\frac{3}{2},\,\dots,\,n-\frac{1}{2}\,,\quad{\rm and}
\mea
\mea
\mea
\!\!{\cal S}({\widetilde {\cal V}}_{k})
&=&\frac{1}{4(k+1)}\left(2k+1,\,
\frac{2k+3}{3},\,
\frac{2k+3}{3},\,
\frac{2k+3}{3}
\right),
\quad
\mea
&\quad&\quad{\rm for}\quad
k=\frac{1}{2},\,\frac{3}{2},\,\dots,\,n-\frac{3}{2}\,.
\mea
\label{HatomShmidts}
\eea
Now we see a remarkable ordering of these subspaces by their spin-orbit
entanglement.  From least entangled
to most entangled, as indicated by their Schmidt strings, we have:
\newpage
\bea
{\cal S}({\cal V}_{1/2})
&=&\Big(1,\,0,\,0,\,0\Big)
\succ\,
{\cal S}({\cal V}_{3/2})
\succ\,
\dots\qquad\qquad
\mea
\mea
\dots \succ
&\!&\!\!\!\!\!\!\!\!\!\!\!
{\cal S}({\cal V}_{n-1/2})
=\frac{1}{4n-2}\Big(2n,\,
\frac{2n-2}{3},\,
\frac{2n-2}{3},\,
\frac{2n-2}{3}
\Big)
\mea
\mea
\mea
\succ\,
{\cal S}_0&=&\Big(\frac{1}{2},\,
\frac{1}{6},\,
\frac{1}{6},\,
\frac{1}{6}
\Big)
\succ\,
\mea
\mea
\mea
{\cal S}({\widetilde {\cal V}}_{n-3/2})
&=&\frac{1}{4n-2}\Big(2n-2,\,
\frac{2n}{3},\,
\frac{2n}{3},\,
\frac{2n}{3}
\Big)
\succ\, \dots
\mea
\mea
\dots\,\succ
&\!&\!\!\!\!\!\!\!\!\!\!\!
{\cal S}({\widetilde {\cal V}}_{3/2})
\succ\,
{\cal S}({\widetilde {\cal V}}_{1/2})
=\Big(\frac{1}{3},\,
\frac{2}{9},\,
\frac{2}{9},\,
\frac{2}{9}
\Big)\,.
\mea
\label{majorizedsubspaces}
\eea
Here the limiting Schmidt string 
${\cal S}_0$ as in (\ref{limitingSchmidt}), is approached from above by ${\cal S}({\cal
V}_{n-1/2})$, and from below by ${\cal S}({\widetilde {\cal V}}_{n-3/2})$, as $n\to \infty$. 

In this  example, 
it seems that the notion
of subspace 
entanglement has to be interpreted as a kind of mean entanglement per basis vector, rather
than a total entanglement.   
Thus the 2-dimensional subspace ${\widetilde{ \cal V}}_{1/2}$, according to this
notion,
has a greater
entanglement than, say, ${\cal V}_{n-1/2}$, although the latter subspace may be of much greater
dimension, containing  many entangled states.  

\section{Concluding remarks}

The notion of subspace entanglement that has been introduced 
has some interesting features.
It is desirable in future work 
to try and establish that it does indeed make good sense in the context of applications to physics,
perhaps along the lines that have been explored \cite{vidal} in the case of state entanglement.
To do that, it may well be necessary to 
relate more closely than we have done here, the entanglement of
individual basis vectors in a subspace, with the notion of subspace entanglement.  

Are there other ways to measure subspace entanglement? Another way 
might be to consider an
arbitrary  orthonormal
basis of the subspace, and to consider the system to be  in a mixed state of those  
basis states, each with probability $1/d$, where $d$ is the subspace dimension.
Then we could associate the entanglement of that mixed state
with the entanglement of the
subspace, using existing measures of entanglement of mixed states \cite{wootters,bennett,vedral,vidal2}.  
The density operator for the mixed state in this case is 
simply a multiple (by $1/d$) of the projection operator onto the subspace, so we would then be 
considering in a different way,  
the entanglement associated with  a projection  operator.  
It should be interesting to explore the connections between these two notions of subspace
entanglement.  

We might also consider the possiblity that the subspace of states associated
with a given quantum system is itself uncertain.  In that situation
it would seem appropriate to consider the
extension of the superoperator density matrix $\RSO$ to the mixed case, with probabilties
$p_1,\,p_2,\,\dots,\,p_N$ associated with different subspaces 
$\Vspace_1,\,
\Vspace_2,\,
\dots,\,
\Vspace_N$. 
Then we would need to extend existing notions of entanglement of mixed states to this new
situation.

We have considered above the coupling of an angular momentum $j$ with a spin $1/2$.  
There is a challenge to calculate the entanglement of irreducible subspaces 
with definite total angular momentum, in the case when two arbitrary angular momenta are
coupled together.  When this is done, it should be possible
to see how the entanglement is related to the values of Clebsch-Gordan
coefficients, in particular.  
This `reduction entanglement'
problem has an obvious extension to representations of 
other groups and algebras, and seems to open up a new aspect of the tensor product reduction
problem in general, including cases involving
infinite-dimensional representations, and cases involving tensor products of more than two
representations.

The concept of entanglement of a vector subspace of a tensor product space seems clearly to be
of mathematical interest.  It is less clear what may be its importance for physics, but we
hope that the examples above are suggestive of important applications.

\end{document}